\documentclass[graybox]{svmult}

\usepackage{mathptmx}       
\usepackage{helvet}         
\usepackage{courier}        
\usepackage{type1cm}        
\usepackage{makeidx}         
\usepackage{graphicx}        
\usepackage{multicol}        
\usepackage[bottom]{footmisc}

\usepackage{epsf}

\makeindex             

\def\simlt{\lower.5ex\hbox{$\; \buildrel < \over \sim \;$}}
\def\simgt{\lower.5ex\hbox{$\; \buildrel > \over \sim \;$}}

\def\be{\begin{equation}}
\def\ee{\end{equation}}
\def\beq{\begin{eqnarray}}
\def\eeq{\end{eqnarray}}
\def\bA{{\,\mathbf A}}
\def\bB{{\,\mathbf B}}
\def\bE{{\,\mathbf E}}
\def\bj{{\,\mathbf j}}

\def\Rmax{R_{\rm max}}

\def\ff{f_\star}

\def\ta{\psi}

\def\uf{u_{\star}}

\def\If{I_{\star}}

\def\BQ{B_Q}

\def\V{{\cal V}}

\def\tev{t_{\rm ev}}

\def\XTE{XTE~J1810-197}
\def\AXP{1E~1547.0-5408}

\def\Epar{E_\parallel}
\def\Ebr{E_{\rm br}}

\def\M{{\cal M}}
  \def\C{C}
\def\nmin{n_{\rm min}}

\def\F{{\cal F}}
\def\D{{\cal D}}

\newbox\grsign \setbox\grsign=\hbox{$>$} \newdimen\grdimen \grdimen=\ht\grsign
\newbox\simlessbox \newbox\simgreatbox \newbox\simpropbox
\setbox\simgreatbox=\hbox{\raise.5ex\hbox{$>$}\llap
     {\lower.5ex\hbox{$\sim$}}}\ht1=\grdimen\dp1=0pt
\setbox\simlessbox=\hbox{\raise.5ex\hbox{$<$}\llap
     {\lower.5ex\hbox{$\sim$}}}\ht2=\grdimen\dp2=0pt
\setbox\simpropbox=\hbox{\raise.5ex\hbox{$\propto$}\llap
     {\lower.5ex\hbox{$\sim$}}}\ht2=\grdimen\dp2=0pt
\def\simgt{\mathrel{\copy\simgreatbox}}
\def\simlt{\mathrel{\copy\simlessbox}}

\begin{document}

\title*{Activated Magnetospheres of Magnetars}
\author{Andrei M. Beloborodov}
\institute{Andrei M. Beloborodov \at 
Physics Department and Columbia Astrophysics Laboratory,
Columbia University, 538  West 120th Street New York, NY 10027,
\email{amb@phys.columbia.edu}
}
%
%
\maketitle

\vspace*{-1.2cm}

\abstract{
Like the solar corona, the external magnetic field of magnetars is twisted 
by surface motions of the star. The twist energy is dissipated over time.
We discuss the theory of this activity and its observational status.
(1) Theory predicts that the magnetosphere tends to untwist in a peculiar 
way: a bundle of electric currents (the ``j-bundle'') is formed 
with a sharp boundary, which shrinks toward the magnetic dipole axis. 
Recent observations of shrinking hot spots on 
magnetars are consistent with this behavior.
(2) Continual discharge fills the j-bundle with $e^\pm$ plasma,
maintaining a nonthermal corona around the neutron star.
The corona outside a few stellar radii strongly interacts with the 
stellar radiation and forms a ``radiatively locked'' outflow with a high 
$e^\pm$ multiplicity. The locked plasma annihilates near the apexes of the 
closed magnetic field lines.
(3) New radiative-transfer simulations suggest a simple mechanism 
that shapes the observed X-ray spectrum from 0.1~keV to 1~MeV: part of the 
thermal X-rays emitted by the neutron star are reflected from the outer 
corona and then upscattered by the inner relativistic outflow in the 
j-bundle, producing a beam of hard X-rays.
}

\section{Introduction}
\label{sec:1}
Term ``magnetars'' was coined for neutron stars with ultrastrong magnetic 
fields $B\sim 10^{14}-10^{15}$~G (Duncan \& Thompson 1992; Paczy\'nski 1992).
There is substantial evidence for ultrastrong fields in soft gamma repeaters 
(SGRs) and anomalous X-ray pulsars (AXPs) (see e.g. reviews by Woods \& 
Thompson 2006; Mereghetti 2008). Alternative models for SGRs and AXPs are 
not discussed here.

The current magnetar catalogue contains 18 objects (13 confirmed and 
5 candidates).\footnote{
      http://www.physics.mcgill.ca/~pulsar/magnetar/main.html
      } 
Host supernova remnants were identified for some of them,
confirming their young ages $t\sim 10^3-10^5$~yr.
The young age, combined with the currently observed number of objects, 
implies that $\simgt 20$\% of all neutron stars  are born as magnetars.

The rotation rates of all observed magnetars are moderate by neutron-star 
standards. Their spin periods $P$ are in a rather narrow range from 2 to 
12~s. On the other hand, their spindown rates $\dot P$ show large temporal 
variations, usually related to the X-ray outbursts. This activity
contrasts with the traditional view of neutron stars as passive 
stellar remnants. Unlike ordinary pulsars, whose activity is associated with 
the open field lines that extend to the light cylinder, the magnetar 
activity must be generated in the {\it closed} magnetosphere.
Its energy output greatly exceeds the rotational energy of the star
and must be fed by magnetic energy. 

A distinct feature of these objects is their huge nonthermal luminosity, 
exceeding the spindown power by 1-3 orders of magnitude. Between the 
outbursts, magnetars display persistent or slowly decaying X-ray emission
with luminosities $L\sim 10^{34}-10^{35}$~erg~s$^{-1}$.
Two peaks are observed in the X-ray spectrum:
one near keV and the other above 100~keV (see e.g. Kuiper et al. 2008). 
The energy fluxes in the two peaks are comparable.
The keV peak is dominated by emission from the hot surface of the 
neutron star. The 100-keV component is clearly nonthermal
and demonstrates the existence of a hot corona around magnetars, 
which provides a large fraction of the total observed luminosity.
For comparison, the corona of the sun radiates 
only $\sim 10^{-6}$ of the solar luminosity.

The observed magnetospheric activity must be generated by motions of 
the star surface. The crust of a magnetar
is stressed by the ultrastrong magnetic field and can yield to the 
stresses through episodic starquakes or a slow plastic flow
(Thompson \& Duncan 1996).
The magnetosphere is anchored in the crust and inevitably 
twisted by the crustal motions, resembling the behavior of the solar corona. 
It becomes non-potential, $\nabla\times\bB\neq 0$, and threaded by 
electric currents 
(Thompson et al. 2002, hereafter TLK02; 
Beloborodov \& Thompson 2007, hereafter BT07). 
The currents can flow only along $\bB$, and the twisted magnetosphere 
remains nearly force-free, $\bj\times\bB=0$.

No quantitative theory has been developed for the crust motion in magnetars.
The theory does not predict the geometry of the magnetosphere and its 
deformations, however, it is possible to predict what happens after a 
twist has been implanted. Part of the magnetosphere becomes 
filled with plasma and gradually dissipates the magnetic energy, creating 
a long-lived luminous corona. The theory of this activity can be tested 
with observations of the post-burst behavior of magnetars. 
Besides the nonthermal emission, 
bright hot spots are observed on magnetars after their bursting activity.
The spots slowly shrink, on a timescale of months to years.
Understanding this behavior may help reconstruct the geometry of 
the implanted twists and perhaps the crust motion that created them. 

The theory is developed in two steps. First, an electrodynamic model is 
formulated to describe the deformed magnetosphere and its evolution (\S~2). 
Second, basic properties of plasma filling the magnetosphere are evaluated 
(\S\S~4 and 5). Observational tests of the theory are discussed in \S\S~3 
and 6.

In this contribution, we focus on the X-ray emission mechanism in magnetars.
The related topics of their spindown behavior and optical/IR/radio emission 
are discussed elsewhere in this volume.

\section{Electrodynamics of untwisting}
\label{sec:3}
Twisted force-free magnetospheres were extensively studied in the context 
of the solar corona. In axial symmetry, they are described by the
Grad-Shafranov equation, and various force-free equilibria can be 
constructed numerically by solving this equation.
A simple example is a self-similarly twisted dipole (Wolfson 1995).
Pavan et al. (2009) studied multipolar self-similar solutions. 
Realistic non-self-similar equilibria do not admit a simple 
description and are less explored, except for the case of weak 
twists (Wolfson \& Low 1992; Beloborodov 2009, hereafter B09). 

The twisted magnetosphere stores additional energy -- the magnetic energy 
of the electric currents maintaining $\nabla\times\bB$.
The magnetosphere tends to dissipate the stored energy and untwists with 
time. Apart from the rare flares, the evolution proceeds slowly 
through a sequence of force-free equilibria.
A key agent in this evolution is the moderate electric field 
$\Epar$ parallel to $\bB$. It has three important functions (BT07): 
(1) maintains the current flow, 
(2) regulates the dissipation rate $\Epar j$ and the observed luminosity, and 
(3) determines the evolution of $\bB$ in the untwisting magnetosphere. 

This section describes the untwisting theory for axisymmetric magnetospheres.
A simple electrodynamic equation governs the evolution of $\bB$.

\subsection{Evolution equation for axisymmetric twist}
Any axisymmetric magnetic field $\bB$ may be expressed in the following form,
\be
\label{eq:B}
  \bB=\frac{\nabla f\times{\mathbf e}_\phi}{2\pi r\sin\theta}
     +B_\phi {\mathbf e}_\phi.
\ee
Here $r,\theta,\phi$ are spherical coordinates, $\phi$ is the angle of 
rotation about the axis of symmetry, and ${\mathbf e}_\phi$ is the unit
vector in the $\phi$ direction. For example, a dipole magnetosphere with 
dipole moment $\mu$ has $f(r,\theta)=2\pi\mu \sin^2\theta/r$ and $B_\phi=0$.
Function $f(r,\theta)$ has the meaning of the magnetic flux through 
the circle $0<\phi<2\pi$ of constant $r$ and $\theta$. 
Note that $f$ is constant along any field line ($\bB\cdot\nabla f=0$) and
by symmetry $f$ is also constant on any 2D ``flux surface'' formed by 
rotating a field line about the symmetry axis. 
Any axisymmetric magnetosphere
can be thought of as a set of nested flux surfaces labeled by $f$.
Note that $f=0$ on the axis. 

If $B_\phi=0$ then the axisymmetric magnetosphere is potential,
$\nabla\times \bB=0$, and carries no electric currents. 
Any axisymmetric starquake is, in essence, a latitude-dependent rotation 
of the crust. It inevitably imparts a twist into the magnetosphere.
Its amplitude is measured by the azimuthal extension of field lines
outside the star,
\be
    \psi(f)=\phi_Q-\phi_P=\int_P^Q \frac{B_\phi}{B}\,\frac{dl}{r\sin\theta},
\ee
where the intergal is taken along the field line between its footprints 
$P$ and $Q$ on the star surface. The appearance of $B_\phi\neq 0$ 
implies a non-zero circulation of $\bB$ along circles of constant $r,\theta$. 
The circulation must be sustained by an electric current 
through the circle, according to Stokes' theorem.

Current can flow only along $\bB$ and must be 
maintained by a longitudinal voltage,
\begin{eqnarray}
\label{eq:Phi_e}
  \Phi_e=\int_P^Q \Epar dl,
\end{eqnarray}
where $\Epar$ is the electric field component parallel to $\bB$. 
Note that both footprints $P$, $Q$ of the field line are ``grounded'' in 
a good conductor --- the neutron star --- and the electrostatic voltage 
between $P$ and $Q$ is negligible. Voltage $\Phi_e$ is {\it inductive}, 
not electrostatic.
It is directly related to the rate of untwisting of the magnetic field,
$\int \Epar dl=c^{-1}\int(\partial A_\parallel/\partial t)\, dl$
where $\bA$ is the magnetic vector potential (BT07).
The gradual dissipation of the magnetospheric twist is, in essence, 
Ohmic dissipation in the electric circuit. The dissipation rate 
is proportional to $\Phi_e$.

\begin{figure}[t]
\sidecaption
\includegraphics[scale=.45]{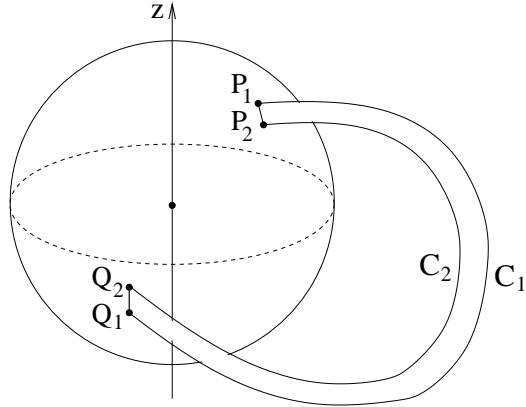}
\caption{
Field line $C_1$ with footprints $P_1$, $Q_1$ on a flux surface $f_1$ 
and its neighbor field line $C_2$ with footprints $P_2$, $Q_2$ on a 
flux surface $f_2$. Closed contour $C$ is formed by connecting $P_1$ 
with $P_2$ and $Q_1$ with $Q_2$ along the star surface.
}
\label{fig:1}       
\end{figure}

How does the twist $\psi(f)$ evolve as a result of Ohmic dissipation?
The electrodynamic equation $\partial \bB/\partial t=-c\nabla\times\bE$
states that electric field determines the evolution of magnetic field.
Its integral version (Faraday's law) can be used to derive the evolution
equation for $\psi$.
Consider two infinitesimally close field lines $C_1$ and $C_2$ on 
different flux surfaces $f_1$ and $f_2=f_1+\delta f$. A closed contour $C$ 
can be formed by connecting the footprints along the stellar surface:
$P_1$ with $P_2$ and $Q_1$ with $Q_2$ (Fig.~1). 
Let $\delta \Phi$ be the magnetic flux through contour $\C$.
The Faraday's law states
\be
\label{eq:far}
  \frac{1}{c}\frac{d\delta\Phi}{d t}=-\oint_{\C} \bE\cdot d{\mathbf l},
  =\int_{C_2} \Epar dl -\int_{C_1} \Epar dl,
\ee
where we neglected the infinitesimal contribution from curves $P_1P_2$
and $Q_1Q_2$ to the line integral $\oint_C$.
Only $B_\phi$ contributes to flux $\delta\Phi$. This can be seen if we
continuously deform $C_1$ and $C_2$ so that their points slide at 
constant $r,\theta$ in the $\phi$-direction until both $C_1$ and $C_2$ 
end up in one meridional plane $\phi=0$. 
In this deformation, $C_1$ and $C_2$ slide along their flux surfaces 
$f_1$ and $f_2$, and hence $\delta\Phi$ remains unchanged by the deformation
(magnetic flux through any part of a flux surface is zero).
Only $B_\phi$ contributes to the magnetic flux through the deformed contour 
in the meriodinal plane. One can show that $\delta\Phi$ is related to the 
twist angle $\psi(f)$ by
\be
\label{eq:ident}
  \delta\Phi=\frac{\psi(f)\,\delta f}{2\pi}.
\ee
This geometrical identity gives a quick way to derive the evolution
equation for $\psi(f,t)$. Combining equations~(\ref{eq:Phi_e}),
(\ref{eq:far}), and (\ref{eq:ident}) we immediately obtain\footnote{
    This evolution equation was derived in B09 for weak twists $\psi\ll 1$.
    As the derivation here shows, the same equation is also valid  
    for strong twists. Equation~(17) in B09 is not accurate for strong 
    twists, it neglects the motion of flux surfaces in the process 
    of untwisting.}
\begin{equation}
\label{eq:evol}
  \left.\frac{\partial\psi}{\partial t}\right|_f
  =2\pi c\frac{\partial\Phi_e}{\partial f}.
\end{equation}
Here we used $d/dt=(\partial/\partial t)_f$ taking into account that 
contour $C$ remains on the same flux surfaces $f$, $f+\delta f$ as
the magnetosphere untwists (the flux surfaces can change their shapes
in the evolving magnetosphere).

Equation~(\ref{eq:evol}) describes the twist evolution 
assuming that the footprints 
of field lines are frozen in the static crust of the star. The untwisting 
of field lines is due to their ``slipping'' in the magnetosphere, 
which changes the connectivity between the footprints so that $\psi$
is reduced.
It is straightforward to generalize the evolution equation for 
magnetospheres with moving footprints,
\be
\label{eq:evol1}
   \frac{\partial\psi}{\partial t}=2\pi c\frac{\partial\Phi_e}{\partial f}
     +\omega(f,t),
\ee
where $\omega=d\phi_Q/dt-d\phi_P/dt$ is the differential angular velocity
of the moving crust.

\subsection{j-bundle}
  Equation~(\ref{eq:evol}) has an interesting feature: the evolution 
of $\psi$ on a given flux surface $f$ is governed by the transverse 
{\it gradient} of the voltage $\partial\Phi_e/\partial f$ rather than its 
value $\Phi_e(f)$. This leads to a paradoxical result: 
 if $\Phi_e(f)=const$ i.e. the voltage is the same for all field lines, 
then the twist must ``freeze'' --- the magnetic configuration does not 
change with time. How can this be reconciled with the continuing Ohmic 
dissipation in the electric circuit $\Phi_e j\neq 0$?
The circuit is ought to dissipate the magnetic (twist) energy, and
the unchanged $\bB$ appears to violate energy conservation.

The paradox is resolved if one notices that $\Phi_e$ cannot be the same
for {\it all} field lines. 
In particular, field lines $f>f_R$ are confined inside the star
where $\Epar\approx 0$. To fix ideas, suppose the star is a perfect 
conductor, which implies $\Phi_e(f)=0$ for $f>f_R$. 
Only field lines extending beyond the stellar surface, i.e. forming the 
magnetosphere, have $\Phi_e\neq 0$, which implies a gradient 
$\partial\Phi_e/\partial f\neq 0$ near the flux surface $f=f_R$.
Then the twist evolution equation~(\ref{eq:evol}) implies that 
$\partial\psi/\partial t< 0$ is maintained near $f=f_R$.
Effectively, the magnetospheric electric currents are ``sucked'' into 
the star through the flux surface $f_R$. This behavior is 
demonstrated by the explicit solutions to the 
evolution equation~(\ref{eq:evol}) in B09.

There is another nontrivial feature of the untwisting process.
Under normal conditions, plasma density in the closed 
magnetosphere\footnote{ 
    The usual estimate for plasma density is $\rho_{\rm GJ}/e$ where 
    $\rho_{\rm GJ}=-\Omega\cdot\bB/2\pi c$ is the charge 
    density induced by rotation of the star with an angular velocity 
    $\Omega$ (Goldreich \& Julian 1969).
    }
is not sufficient to conduct interesting electric currents.
Plasma in the twisted magnetosphere is supplied by a continual discharge, 
which has a threshold voltage $\Phi_e=\V(f)\sim 10^9$~V (BT07). 
The voltage is well described by a step function (see \S~3 for details),
\begin{eqnarray}
\label{eq:V}
  \Phi_e=\left\{\begin{array}{ll}
                   \V & j>0 \\
                   0  & j=0 \\
                   -\V & j<0 \\
                \end{array}
         \right.
\end{eqnarray}
The threshold nature of the discharge has a rather peculiar implication.
Two distinct regions immediately form in an untwisting magnetosphere: 
``cavity'' near the flux surface $f_R$ and ``j-bundle'' where $j\neq 0$ 
(see B09 and Fig.~2). The two regions are separated by 
a sharp boundary along a flux surface $\ff$. The untwisting process is 
the slow motion of the boundary $\ff$ that gradually erases the remaining 
electric currents. The erased 
magnetospheric currents end up closed below the stellar surface.

If the discharge voltage $\V(f)=const$, the twist freezes in 
the j-bundle $f<\ff$ and passively waits for the front $\ff$ to come.
Realistically, $\V(f)\neq const$ and $d\V/df>0$ is possible.
Then equation~(\ref{eq:evol}) gives $\partial\psi/\partial t>0$, 
i.e. the twist {\it grows} inside the j-bundle while waiting 
for the front to come. 
In all cases, the magnetic energy of the shrinking j-bundle decreases
with rate equal to the rate of Ohmic dissipation.

\begin{figure}[t]
\hspace{-0.3cm}
\epsfxsize=12.1cm \epsfbox{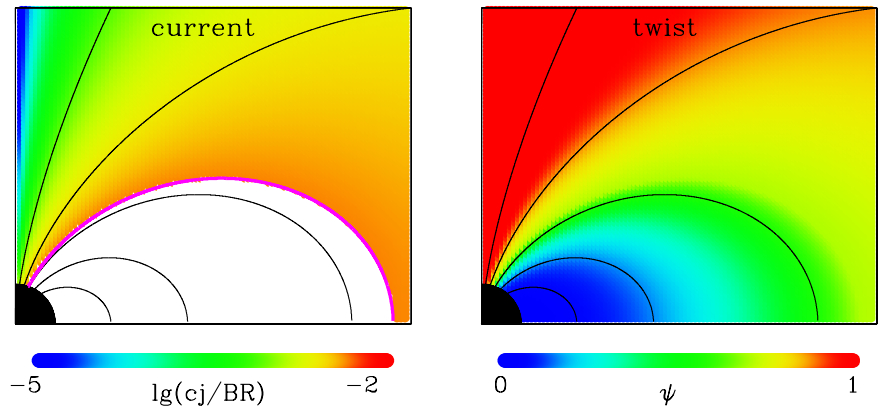} 
\vspace{-0.3cm}
\caption{
Snapshot of the magnetospheric evolution. A global twist with a moderate 
$\psi=0.2$ was implanted into the magnetosphere at $t=0$, and the 
snapshot shows the magnetosphere at $t\sim 1$~yr. 
Details of the calculations are described in B09. 
The plane of the figure is the poloidal cross section of the magnetosphere.
The black curves are the poloidal magnetic field lines; they are symmetric 
about the vertical axis and the equatorial plane. 
{\bf Left panel:} current density. The cavity is shown in white ($j=0$); 
its boundary $\ff$ is marked by the magenta curve.
The cavity expands with time, and the j-bundle shrinks toward the vertical
axis. {\bf Right panel:} twist amplitude $\psi$ at the same time. 
The neutron star is shown by the black circle.
}
\label{fig:1}       
\end{figure}

The behavior of untwisting magnetospheres is illustrated 
in B09 by solving the electrodynamic equation~(\ref{eq:evol}) for a 
concrete configuration: a centered dipole with a weak twist $\psi<1$.
The ``cavity + j-bundle'' structure forms immediately, and then the 
cavity gradually expands, while the twist angle $\psi$ inside the j-bundle 
grows linearly with time. This growth leads to $\psi>1$ where the model 
breaks. General considerations (e.g. Uzdensky 2002) show that
the magnetosphere with $\psi>\psi_{\rm cr}={\cal O}(1)$ must
lose equilibrium and restructure itself.
Two possible types of restructuring are known:
(1) Partial opening of field lines, which can lead to a configuration 
with lower energy (Wolfson \& Low 1992). In an axisymmetric magnetosphere, 
this change may occur without breaking the symmetry. (2) Kink instability, 
which breaks the axial symmetry.
Then an asymmetric plasmoid must be ejected by the magnetosphere 
and the twist amplitude must be reduced. The exact threshold 
for the kink, the energy of the ejected plasmoid, and the new equilibrium 
configuration following the kink are unknown. 
One can expect that the amplitude of the growing twist is regulated 
by the instability so that it remains close to the critical value 
$\psi\sim 1$.

The j-bundle and its footprints can be a bright source of radiation,
with luminosity equal to the rate of Ohmic dissipation in the bundle. 
If the footprint of the j-bundle experiences
heating by the bombarding magnetospheric particles (BT07) then 
$\ff$ defines the boundary of a hot spot on the star $\theta<\theta_\star$. 
The area of this spot is given by
\be
\label{eq:area}
   A\approx \pi(R\sin\theta_\star)^2=\pi R^2\uf.
\ee
Here we used the dimensionless variable $u=f/f_R$ and approximated 
the poloidal structure of the magnetosphere by 
the dipole field.\footnote{Dipole 
    field has $f/2\pi=\mu\sin^2\theta/r=\mu/\Rmax$, where $\mu$ 
    is the magnetic dipole moment. Then $u=R/\Rmax=\sin^2\theta_1$ where 
    $\theta_1$ is the polar angle of the field-line footprint on the 
    stellar surface.}
As the cavity expands in an untwisting magnetosphere (Fig.~2), the spot
$\theta<\theta_\star$ shrinks.

The simplest model with $\V(u)=const$ gives the following 
estimate for the j-bundle luminosity (B09),
\be
\label{eq:lum}
   L=\V\If\approx \frac{c\mu}{4R^2}\,\ta\,\V\,\uf^2
   \approx 1.3\times 10^{34}\,\V_9\,B_{14}\,R_6\,\ta\,
     \left(\frac{\uf}{0.1}\right)^2\;{\rm erg~s}^{-1},
\ee
where $\V_9\equiv\V/10^9$~V and $B_{14}=B_{\rm pole}/10^{14}$~G.
The evolution timescale of the luminosity is given by
\be
\label{eq:tev}
  \tev=-\frac{L}{dL/dt}\approx \frac{\mu\uf}{cR\V}
      \approx 1.5\,\V_9^{-1}\,B_{14}\,R_6^2\,\ta\,
      \left(\frac{\uf}{0.1}\right)\;{\rm yr}.
\ee
Similar calculations have been done for the more detailed model where 
voltage $\V$ is different on different field lines (B09).
The theoretical expectations are compared with observations in the next 
section.

\section{Transient magnetars}
\label{sec:3}
Recent observations revealed that many 
magnetars spend a significant fraction of time in a quiescent state
with a low luminosity $L\sim 10^{33}-10^{34}$~erg~s$^{-1}$ and remain
unknown until they produce an X-ray outburst. The post-outburst decay 
of the luminosity back to the quiescent level is monitored in 
these ``transient magnetars.'' The data can be compared with 
the theoretical model of untwisting magnetosphere.

\subsection{Magnetospheric activity or deep crustal heating?}
The mechanism of the magnetar bursts 
(e.g. Thompson \& Duncan 1995; Woods \& Thompson 2006) is not settled yet. 
It may be associated with starquakes. It may also be associated with sudden 
reconnection events in the magnetosphere that is slowly deformed by the 
plastic crustal flow (Lyutikov 2003). In either case, a strong 
and brief magnetospheric dissipation is invoked to explain the burst emission. 

Following the burst, a bright afterglow 
is observed, which continues to decay for months to years.
The blackbody area of this emission is much smaller than the surface area
of the neutron star, and it can be associated with the hot footprint of 
the j-bundle in the untwisting magnetosphere. This model relies on 
the external heating of the stellar surface by the energetic particles 
 accelerated in the j-bundle.

An alternative to external heating was considered by Lyubarsky et al.
(2002). Their model assumed that during the burst a significant heat is  
deposited inside a deep region of the stellar crust. The subsequent passive
cooling of this region due to heat diffusion through the crust can produce 
a bright thermal afterglow. Its expected spectrum is quasi-blackbody.

Observations of several transient magnetars in the past few years 
give enough information to judge which of the two heating mechanisms 
--- external or internal --- is responsible for the long-lived afterglow.
Two features point to external heating.

\begin{itemize}

\item
{\it Nonthermal X-ray spectrum.} 
The X-ray emission shows significant deviations from blackbody (BB)
and is usually fitted by BB + power-law model.
A high-energy X-ray component was discovered in the transient magnetar 
1E~1547.0-5408 (Enoto et al. 2010a). 
Its luminosity is comparable to the luminosity of the blackbody component,
and the overall spectrum of 1E~1547.0-5408 was similar 
to the spectra of magnetars with persistent coronal activity.
The energy dependence of pulse profiles resembled that
of persistent emission in 4U~0142+61 and 1RXS~J170849-400910
(den Hartog et al. 2008a,b), suggesting a similar magnetospheric emission 
mechanism.
In another transient magnetar SGR~0501+4516, bright 100-keV emission was 
reported at $t=4$~d after the outburst (Enoto et al. 2010b), with a similar
spectrum. 
\medskip

\item
{\it Shrinking area of the hot spot.} 
The emission area of the blackbody component $A$ decreases following the 
outburst together with its luminosity $L$. 
The transient blackbody emission evolves along tracks in the $A$-$L$ plane 
as seen in Figure~3.
The tracks are inconsistent with the model of impulsively heated deep
crust, as cooling due to heat diffusion would give 
a hot spot that tends to spread with time, not shrink.
The shrinking spot is consistent with 
the untwisting
magnetosphere model (\S~2.2). The theoretical relation between $A$ and $L$
(obtained from eqs.~\ref{eq:area} and \ref{eq:lum}, see caption to Fig.~3)
appears to agree with the data.
Caution should be taken though when interpreting the details in Figure~3, as 
there may be significant systematic errors in observed $A$ --- it depends on 
the details of spectral fits. Besides, the distance $d$ is poorly known for 
many magnetars. Note, however, that the shape of the observed track in the 
$A$-$L$ plane does not depend on $d$, because both $A$ and $L$ scale as $d^2$.

\end{itemize}

\begin{figure}[t]
  \begin{center}
    \begin{minipage}[t]{0.58\linewidth}
\epsfxsize=8.0cm
 \raisebox{-8.0cm}{\epsfbox{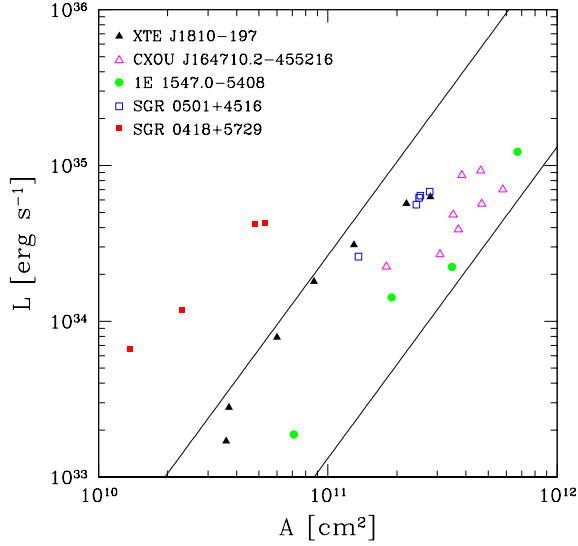}}
    \end{minipage}\hfill
    \begin{minipage}[t]{0.33\linewidth}
\vspace*{0.5cm}
\caption{\small
Area $A$ and luminosity $L$ of the hot spots observed on five transient 
magnetars at different times following their outbursts; see text for 
references and discussion of the individual objects.
One can see from the figure that
the spot area decreases together with its luminosity.
The j-bundle model is shown by the strip between the two lines.
It predicts the relation
$L=1.3\times 10^{33}\,B_{14}\,R_6^{-3}\,\V_9\,\psi\,A_{11}^2$~erg~s$^{-1}$.
$B_{14}\,R_6^{-3}\,\V_9\,\psi=1$ on the lower line and 20
 on the upper line.
}
\end{minipage}
\end{center}
\end{figure}

\noindent
These observations do not rule out deep crustal heating during the 
burst. The internal heating may dominate the thermal afterglow 
during first few days following the burst. 
After a few days, the observed light curve significantly flattens
(Woods \& Thompson 2006), suggesting a possible transition from 
internal to external heating.

\subsection{Individual objects}

{\bf XTE~J1810-197} is a canonical transient magnetar.
It has spin period $P=5.54$~s and magnetic dipole moment 
$\mu\approx 1.5\times 10^{32}$~G~cm$^3$, which corresponds to the surface 
field $B\approx 3\times 10^{14}$~G. An X-ray outburst was detected from 
this object in January 2004 (Ibrahim et al. 2004) and then
its luminosity approximately followed an exponential decay on
a timescale of 233 days for 3 years (Gotthelf \& Halpern 2007).
All data points for this object in Figure~3 were taken during 
the afterglow of this outburst. Remarkably,
the source switched on as a bright radio pulsar (Camilo et al. 2007). 
The radio observations provided an estimate for the distance $d=3.3$~kpc 
(which is used in Fig.~3) and accurate measurements of the spindown rate.
In a few years, the object returned to its quiescent (pre-outburst) state:
the radio pulsations and the spindown enhancement gradually disappeared,
and the X-ray luminosity decreased to its quiescent level.

\begin{figure}[t]
  \begin{center}
    \begin{minipage}[t]{0.58\linewidth}
\epsfxsize=9.5cm
 \raisebox{-9.5cm}{\epsfbox{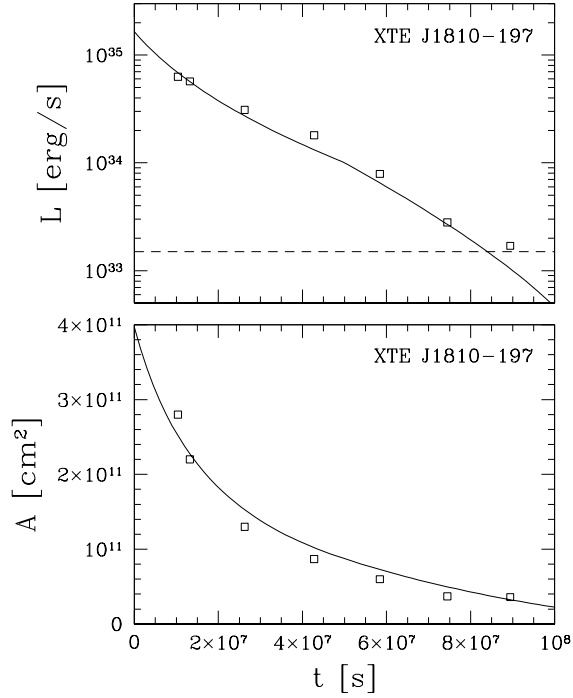}}
    \end{minipage}\hfill
    \begin{minipage}[t]{0.33\linewidth}
\caption{\small
Comparison of the model with the observed evolution of the area $A$ and 
luminosity $L$ of the hot spot after the outburst in \XTE.
The data ({\it open squares}) are from Gotthelf \& Halpern (2007).
{\it Dashed line} shows the object luminosity in quiescence.
{\it Solid curves} show the proposed theoretical model.
The starquake occurred in the region $u<u_0=0.15$ with amplitude $\ta_0=0.5$.
The discharge voltage is assumed to drop linearly from $5.5$~GeV at $u=0.15$ 
to 1~GeV at $u=0$. (From B09.)
}
\end{minipage}
\end{center}
\end{figure}

These observations show that the magnetosphere of \XTE~ changed in the 
outburst. Its footpoints must have moved, imparting a twist in the 
magnetosphere. The data give significant hints about the twist geometry 
and evolution:

(1) The change in spindown rate suggests that
the open field-line bundle was strongly affected, and hence the
twist reached large amplitudes $\ta\simgt 1$ near the magnetic dipole axis.
Recall also that $\ta$ is limited to $\ta_{\rm cr}={\cal O}(1)$ by the MHD 
instability.

(2) The luminosity was observed to decrease from
$\sim 10^{35}$~erg/s to $\sim 10^{33}$~erg/s.
The theoretically expected luminosity from a global twist with
$\ta\sim 1$ would be much higher: $L\sim 10^{37}\V_9$~erg/s
(eq.~\ref{eq:lum}).
We conclude that the twisted region was small: the current-carrying field 
lines formed a narrow bundle emerging from a small spot on the star surface.

(3) The spot was discovered:
a hot blackbody component with a small emission area was
found in the X-ray spectrum following the outburst (Gotthelf \& Halpern 2007;
Perna \& Gotthelf 2008; Bernardini et al. 2009). Its emission area shrank 
with time until the spot became undetectable (Fig.~4).

(4) The X-ray and radio pulse profiles had almost simultaneous peaks,
consistent with the X-ray emitting spot being on the magnetic dipole axis
(Camilo et al. 2007).

 \medskip

This behavior is consistent with the untwisting magnetosphere model
described in \S~2. 
The theoretical estimates for $A$, $L$, and $\tev$ (eqs.~\ref{eq:area},
\ref{eq:lum}, \ref{eq:tev}) are in remarkable agreement with the data.
Figure~4 compares the detailed model with observations of \XTE.
The data constrain the magnetospheric voltage $\V$ to be between 1 and 
6 GeV, which is close to its expected theoretical value (\S~4).
\bigskip

{\bf 1E~1547.0-5408} is known as the magnetar with the shortest 
spin period $P=2$~s. Its estimated magnetic field is 
$B\sim 2\times 10^{14}$~G. The object shares many features of \XTE. 
A similar dynamic hot spot was detected (Halpern et al. 2008).
The data points for \AXP~ in Figure~3 assume $d=4$~kpc following Tiengo et al. 
(2010). One data point shows a pre-outburst state 
and two points show the decay after the outburst in 2007. 
The spot size was decreasing, however later the 
luminosity and the spot grew again.
In October 2008 new burst activity was detected, the emission area 
quickly increased to $A\approx 10^{12}$~cm$^2$ and remained 
approximately constant for 20 days (Israel et al. 2010).
A similar luminosity and area were observed after the outburst in January 
2009 (Enoto et al. 2010a); this gave another data point in Figure~3.
The activity level of 1E~1547.0-5408 is clearly higher than that of \XTE.
Apparently, many deformation events occurred in its magnetosphere in the 
past few years, and it may be hard to see the long-timescale untwisting 
behavior because it is interrupted by new bursting activity. 
\medskip

{\bf CXOU J164710.2-455216} is a transient magnetar with spin period
$P=10.6$~s and magnetic field $B\sim 9\times 10^{13}$~G.
Its outburst in September 2006 was accompanied by the increase of 
persistent luminosity from $\sim 10^{33}$~erg~$s^{-1}$ to 
$\sim 10^{35}$~erg~$s^{-1}$ (Muno et al. 2007).
The subsequent slow decay of emission was observed for about 150 days 
(Woods et al. 2010). All eight data points for this object in Figure~3
are based on these observations; distance $d=5$~kpc was assumed following 
Muno et al. (2007). The decay timescale is comparable to that in \XTE.
\medskip

{\bf SGR~0501+4516} was discovered due to its bursting activity in August 
2008 and then its post-burst behavior was observed 
(Rea et al. 2009). The results of these observations are shown by the 
five data points in Figure~3, assuming $d=5$~kpc. 
The object has spin period $P=5.7$~s and 
magnetic field $B\sim 2\times 10^{14}$~G.
Its post-outburst track in Figure~3 is consistent with a shrinking hot 
spot on the star. The track is close to that of \XTE. The luminosity decayed 
on a timescale of $\sim 24$~days, about 10 times faster than in \XTE. 
This behavior is consistent with the j-bundle model if 
the voltage $\V$ in SGR~0501+4516 is $\sim 3$ times higher and the twist 
amplitude $\psi$ is $\sim 3$ times smaller than the corresponding parameters 
in \XTE. 
This difference in parameters may not affect much the track in the 
$A$-$L$ plane since $L\propto \psi\,\V$ (eq.~\ref{eq:lum}). At the same
time, it explains the short evolution timescale 
since $\tev\propto \psi^{-1}\V$ (eq.~\ref{eq:tev}).
\medskip

{\bf SGR~0418+5729} was discovered due to its outburst in 2009 and its 
post-outburst behavior was followed for 160 days (Esposito et al. 2010). 
Its track in the $A$-$L$ plane is shown in Figure~3 assuming a distance
$d=5$~kpc. The object has spin period $P=9.1$~s and magnetic field 
$B\simlt 3\times 10^{13}$~G. The low magnetic field (by magnetar standards)
makes this object special, which could explain the offset of its track 
from the tracks of other transient magnetars. Note also that the 
track location would be different for different $d$. A large distance 
$d\sim 15$~kpc would imply $\sim 10$ times larger $L$ and $A$, and would 
place the track of SGR~0418+5729 into the strip between the two lines in 
Figure~3. The timescale of the post-burst decay is a few months 
(the luminosity decreased by a factor of 6 in four months).
\medskip

{\bf SGR~1627-41} is another transient magnetar whose 
spin period and magnetic field are close to those of 1E~1547.0-5408.
The object displayed bursting 
activity in 1998 and then its luminosity gradually decayed toward the 
quiescent state with no new bursts detected until 2008 
(Kouveliotou et al. 2003; Mereghetti et al. 2006; Esposito et al. 2008).
The luminosity decayed in $\sim 3-5$-years, with a light curve 
similar to the post-outburst behavior of \XTE. The published spectral 
analysis provides almost no information on the blackbody emission area 
in this object except for one XMM observation when the source luminosity 
was only $4\times 10^{33}$~erg~$^{-1}$ (Mereghetti et al. 2006). 
\medskip

A large fraction of observed magnetars are persistent bright 
sources with luminosities $L\sim 10^{35}$~erg~s$^{-1}$. 
The 100-keV emission component was observed in several of these
``persistent magnetars,'' 
indicating quasi-steady magnetospheric activity (Kuiper et al. 2008). 
Their blackbody components have emission areas $A\simgt 10^{12}$~cm$^2$
(e.g. Perna et al. 2001), significantly larger than in transient magnetars 
with similar luminosities $L\sim 10^{35}$~erg$^{-1}$.
The reason for this difference is presently unclear.
Quasi-steady magnetospheric emission 
is expected if the twist decay timescale $\tev$ is longer than the 
timescale of the crust motion imparting the twist.
 Note that $\tev$ may be long if the voltage along the twisted field 
lines is somewhat below $10^9$~V and/or the magnetic flux in the j-bundle 
is large (i.e. $\uf$ is large, cf. eq.~\ref{eq:tev}). 
Moreover, it is possible that the j-bundle is persistently deformed by 
the plastic flow of the crust, and a quasi-steady twist is 
maintained near the threshold of MHD stability, $\psi\sim 1$.

The luminosity of persistent magnetars $L\sim 10^{35}$~erg~s
(e.g. Durant \& van Kerkwijk 2006) is consistent
with a twist created by rotation of a small cap on the star, with area 
$A\simlt 10^{12}$~cm (see eq.~\ref{eq:lum} and B09). Apparently, the 
footprints of the j-bundle in persistent magnetars are not strongly 
heated, as no hot spots with small $A$ were detected in these objects.

Persistent magnetars also display sudden increases and gradual decays of 
luminosity. 
For example, the outburst of SGR~1900+14 in 1998 led to a rise in its 
persistent luminosity, and in a few months it decayed back to the 
pre-outburst level (Woods et al. 2001). Similar behavior was observed 
in AXPs, for instance in 1E~2259+586 (Woods et al. 2004)
and 4U~0142+61 (Gavriil et al. 2009; Gonzalez et al. 2010).
The rise in luminosity and its subsequent
decay was reported for 1E~1048.1−5937 (Dib et al. 2009).
The transient behavior in these objects is not necessarily associated with 
the same j-bundle that creates their persistent activity.
The additional energy may be released on different 
field lines and after their untwisting the object returns to the 
pre-outburst persistent state. The timescale for this additional activity 
depends on the voltage, twist amplitude, and magnetic 
flux of the transiently deformed field lines. 
The variations in $\tev$ may be similar to those in transient magnetars
-- at least a factor of 10.

Observations of PSR~J1846-0258 suggest that the transient activity 
of the closed magnetosphere may also occur in rotation-powered pulsars 
(Gavriil et al. 2008).


\section{Pair creation}
%
%
Magnetospheric activity manifests itself through relativistic particles 
that can bombard the stellar surface and also generate nonthermal emission.
We now discuss the origin of these particles, their energies, number, 
and circulation in the magnetosphere.

Particles are accelerated along the magnetic field lines by the 
longitudinal voltage $\Phi_e=\int \Epar dl$. The particles can resonantly 
scatter the stellar photons of energy $\hbar\omega$ once they reach the 
Lorentz factor 
\be
\label{eq:res}
  \gamma\approx 10^3 B_{15}\left(\frac{\hbar\omega}{10 \rm ~keV}\right)^{-1},
\ee
which requires $\Phi_e\sim 10^9\,B_{15}\,(\hbar\omega/10{\rm ~keV})^{-1}$~V.
The scattered photons are boosted in energy by the factor of $\sim \gamma^2$.
These high-energy photons convert to $e^\pm$ pairs off the magnetic field,
creating more particles.

A similar process of $e^\pm$ creation operates in the polar-cap discharge 
of ordinary pulsars, but in a different mode. In ordinary pulsars, the 
high-energy photons convert to $e^\pm$ with a delay. The scattered
photon initially moves nearly parallel to $\bB$ and converts to $e^\pm$ 
only when it propagates a sufficient distance where its angle 
$\theta_\gamma$ with respect to $\bB$ increases so that the threshold 
condition for conversion is satisfied. This delay leads to the large 
unscreened voltage in the polar-cap models for pulsars.

In contrast, the magnetic field of magnetars is so strong that pair 
creation can occur {\it immediately} following resonant scattering
(BT07). The energy of the scattered photon is related to its angle 
$\theta_\gamma$ by
\be
  E(\theta_\gamma)=\frac{E_B}{\sin^2\theta_\gamma}
   \left[ 1- \left(\cos^2\theta_\gamma
       +\frac{m_e^2c^4}{E_B^2}\sin^2\theta_\gamma\right)^{1/2}\right],
\ee
where $E_B=(2B/\BQ+1)^{1/2} m_ec^2$ is the energy of the first Landau 
level and $\BQ=m_e^2c^3/\hbar e\approx 4.4\times 10^{13}$~G. 
The scattered photon may immediately be above the threshold for conversion,
$E>E_{\rm thr}=2m_ec^2/\sin\theta_\gamma$, if $B>4\BQ$.
This suggests that $e^\pm$ discharge in magnetars screens $\Epar$ more 
efficiently and buffers the voltage growth.

\bigskip

\noindent 
{\bf Pair creation on field lines with apexes $\Rmax\simlt 2R$}
\medskip

\noindent
The discharge on closed field lines can be explored using a direct 
numerical experiment where plasma is represented by a large number 
of individual particles.
The existing numerical simulations (BT07) describe the discharge 
on field lines that extend to a moderate radius
$\Rmax\simlt 2R$, where $R$ is the radius of the neutron star.
The magnetic field is ultrastrong everywhere along such field lines,
$B\gg\BQ$, and resonant scattering events may be effectively treated 
as events of pair creation -- a significant fraction of scattered
photons immediately convert to $e^\pm$.

The simulations demonstrate that voltage and pair creation self-organize 
so that a particle scatters on average $\sim 1$ photon as it travels 
through the electric circuit, maintaining the near-critical multiplicity 
of pair creation $\M\sim 1$.
This criticality condition requires voltage 
$\V\sim 10^9$~V, which accelerates $e^\pm$ particles to Lorentz factors 
$\gamma\sim 10^3$.
The electric circuit operates as a {\it global} discharge,
in the sense that the accelerating voltage is distributed along the 
entire field line between its footprints on the star. It
is quite different from the localized ``gap'' that is often pictured
above polar caps in pulsars.

The discharge fluctuates on the light-crossing timescale $\sim R/c$ and 
persists in the state of self-organized criticality. The behavior of the 
circuit resembles a continually repeating lightning: voltage between the 
footprints of the field line quasi-periodically builds up and discharges 
through enhanced production of charges. The average plasma density in the 
circuit $n$ is close to the minimum density $\nmin=j/ec$, as required by 
the criticality condition $\M=n/\nmin \sim 1$. 

As discussed in \S~2.2, the currents tend to be quickly
erased on field lines with small $\Rmax$. It may be that this 
part of the magnetospheres of observed magnetars is potential, 
$\nabla\times\bB=0$, with no discharge.
Then the observed activity is associated with currents on field lines
with large $\Rmax$, i.e. extending far from the star. 

\bigskip

\noindent
{\bf Pair creation on field lines with apexes $\Rmax\gg R$}

\medskip

\noindent
The discharge on extended field lines is likely to have a similar 
threshold voltage $\V\sim 10^9$~V, because the conversion of upscattered 
photons to $e^\pm$ is efficient near the footprints where $B\gg\BQ$.
In this zone, particles are able to resonantly scatter soft X-rays once 
they are accelerated to $\gamma\sim 10^3$ (cf. eq.~\ref{eq:res}), 
which requires voltage $\sim 10^9$~V. Further growth of voltage should be 
stopped as it would cause the excessive creation of $e^\pm$ moving in 
both directions, toward and away from the star, leading to efficient 
screening of $\Epar$.
  
The discharge is expected to occur near the star, and some of the created 
particles outflow to $r\gg R$ along the extended field lines. The rate of 
resonant scattering by a relativistic particle {\it increases} as it moves 
from $B\gg \BQ$ to $B\simlt \BQ$. The particle scatters many more photons,
because the resonance condition shifts toward photons of lower energy
$\hbar\omega_{\rm res}\propto B$ whose number density is larger.
Note also that the effective cross section for resonant scattering, 
$\sigma_{\rm res}=2\pi^2 r_ec/\omega_{\rm res}$ increases as $B^{-1}$. 
Photons scattered in the region $B\simgt 10^{12}$~G convert to $e^\pm$,
which scatter more photons.

In essence, the particles outflowing from the discharge zone lose energy 
to scattering as they enter weaker fields, and this energy is transformed 
to new generations of $e^\pm$. A similar cascade above polar caps of 
pulsars was studied recently by Medin \& Lai (2010). As a result, the $e^\pm$ 
multiplicity of the outflow increases from $\M\sim 1$ to $\M\gg 1$.\footnote{
     A steady relativistic outflow without pair creation would have 
     $\M=const$ along the field lines. This follows from conservation of 
     magnetic flux, charge, and particle number, which give
     $n/B=const$, $j/B=const$, and $n/j=const$ along a field line.}
It means no charge starvation in the outer corona --- there are plenty of 
charges to conduct the current demanded by the twisted magnetic field.

\section{Plasma circulation in the magnetosphere}
\label{sec:3}
Dynamics of the $e^\pm$ flow in the outer corona ($r\gg R$) is 
strongly influenced by resonant scattering, which exerts a strong force 
$\F$ on the particles along the magnetic field lines. 
$\F$ vanishes only if the particle has the ``saturation momentum'' 
$p_\star$ such that the radiation 
flux measured in the rest frame of the particle is perpendicular to $\bB$. 
In a weakly twisted magnetosphere with approximately dipole magnetic 
field exposed to central radiation $p_\star$ is given by\footnote{
    This expression is valid in the region where $1-B_r/B>(R/r)^2$. 
    In this region, stellar radiation may be approximated as a central 
    flow of photons, neglecting the angular size of the star $\sim R/r$.}
\be
\label{eq:sat}
    p_\star(r,\theta)=\frac{2\cos\theta}{\sin\theta}
\ee 
(momentum in units of $m_ec$).
The radiative force always pushes the particle toward $p=p_\star$. 
This effect may be measured by the dimensionless ``drag coefficient,'' 
\be
   \D\equiv \frac{r\F}{p\,m_ec^2}.
\ee
Momentum $p_\star$ is a strong attractor in the sense that deviations 
$p-p_\star$ generate $\D\gg 1$ in the outer corona. 

Consider a steady $e^\pm$ outflow in the outer region where no new 
pairs are produced. The outflow satisfies two conditions:
(a) it is nearly neutral, $n_+\approx n_-$, and
(b) it carries the electric current $j$ demanded by $\nabla\times\bB$.
Electric field 
$\Epar$ immediately reacts to violations of these conditions and enforces 
them. From (a) and (b) one can derive the relation between the velocities 
$\beta_\pm$ of $e^\pm$ (details are given in Beloborodov 2010),
\be
\label{eq:j}
    1-\frac{\beta_-}{\beta_+}=\frac{2}{\M+1}.
\ee
Here ``$+$'' corresponds to positrons, which carry current $j_+>0$
and ``$-$'' corresponds to electrons, which carry $j_-<0$;
we assume a positive net current $j=j_++j_-$ for definiteness.
The multiplicity is defined by $\M=(j_++|j_-|)/j$.
An electric field $\Epar$ must be generated in the outflow to 
sustain the condition~(\ref{eq:j}) against the radiative drag that tends 
to equalize $\beta_-$ and $\beta_+$ at $\beta_\star$. This electric 
field is generated by a small deviation from neutrality, 
$\delta n=n_+-n_-\ll n_+$.

The two-fluid dynamics of the outflow is governed by equations,
\begin{eqnarray}
\label{eq:dyn}
    m_ec^2\frac{d\gamma_+}{dl}=\F(\gamma_+)+e\Epar, \qquad
    m_ec^2\frac{d\gamma_-}{dl}=\F(\gamma_-)-e\Epar,
\end{eqnarray}
where $l$ is the length measured along the magnetic field line.
Since $\gamma_-$ and $\gamma_+$ are not independent --- they are related by 
condition~(\ref{eq:j}) --- it is sufficient to solve one dynamic equation, 
e.g. for $\gamma_+$ (and use the dynamic equation for $\gamma_-$ to exclude 
$\Epar$). Straightforward algebra gives,
\be
  m_ec^2\frac{d\gamma_+}{dl}=\frac{\F(\gamma_+)+\F(\gamma_-)}
       {1+d\gamma_-/d\gamma_+}, \qquad 
   \frac{d\gamma_-}{d\gamma_+}=\left(\frac{\M-1}{\M+1}\right)^2
               \left(\frac{\gamma_-}{\gamma_+}\right)^3.
\ee
In the region of strong drag, $|\D|\gg 1$, 
one finds $\F(\gamma_+)\approx -\F(\gamma_-)\approx-e\Epar$.
Here both $e^+$ and $e^-$ components of the outflow have velocities that are
``locked'' by the balance of two strong forces --- radiative and electric.

\begin{figure}[t]
\hspace{-0.6cm}
$\begin{array}{cc}
 \epsfxsize=7.2cm \epsfbox{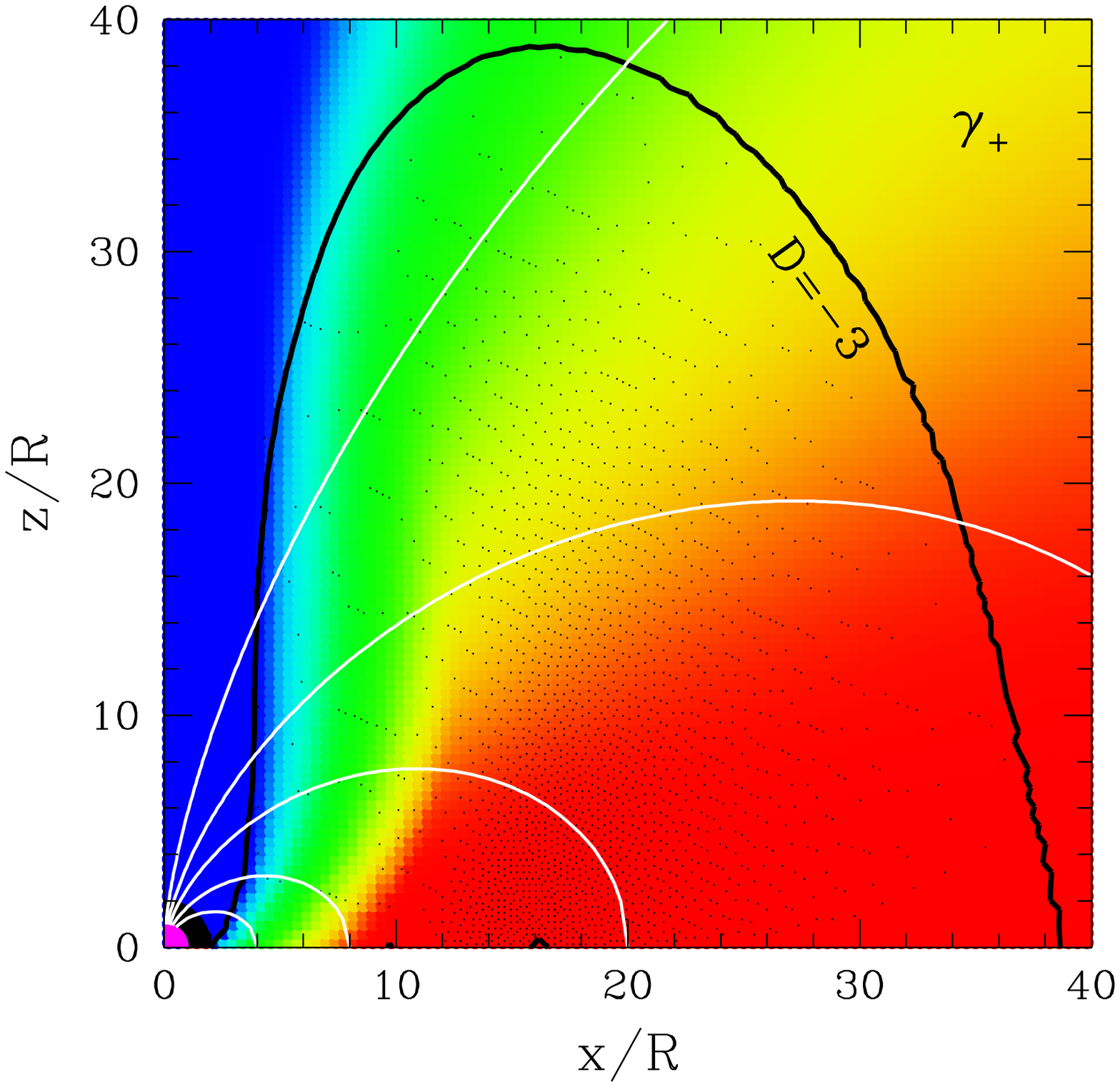} & 
\hspace{-1.6cm} \epsfxsize=7.2cm \epsfbox{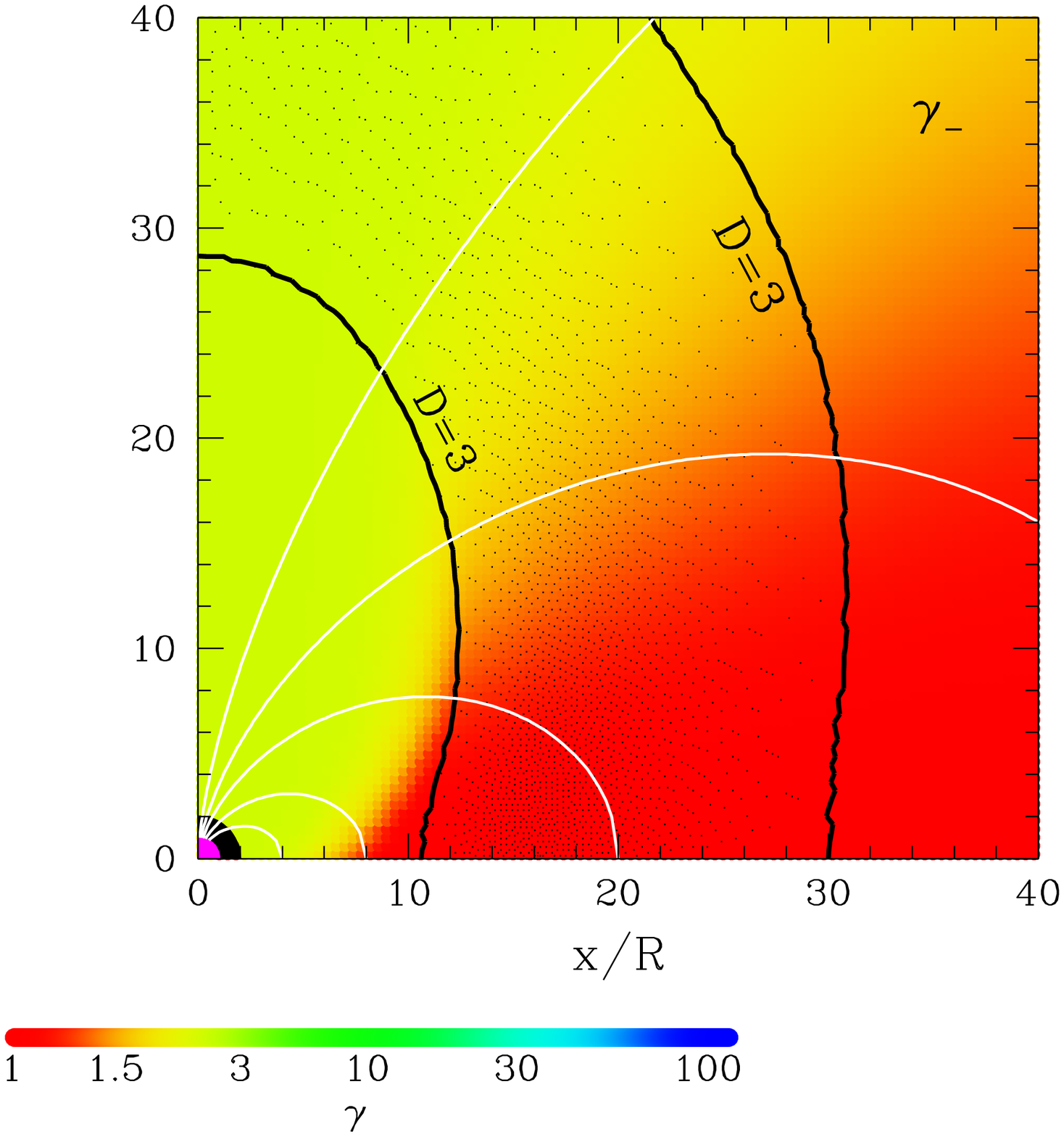}
\end{array}$
\caption{
Lorentz factors $\gamma_+$ (left panel) and $\gamma_-$ (right panel)
in the two-fluid model of outer corona. ~In this example, the electric 
current is carried by the $e^\pm$ outflow of a fixed multiplicity $\M=50$.
The plasma is injected 
at radius $r=2R$ and outflows along the magnetic field lines (white curves). 
The flow is illuminated by the star with temperature $kT=0.5$~keV 
(magenta circle at the origin), which exerts 
radiative forces $\F(\gamma_\pm)$ on the $e^\pm$ plasma.
The Lorentz factors $\gamma_+$ and $\gamma_-$ change as the 
flow enters the drag-dominated region $|\D|\gg 1$. The region $|\D|>3$ 
is shown by the thick black curve and shadowed in black.
$\D<0$ for positrons ($\gamma_+>\gamma_\star$) and $\D>0$ for electrons
($\gamma_-<\gamma_\star$).
~The radiative drag stops the plasma in the equatorial plane
outside $\sim 8R$. A nearly dipole magnetic field (weakly 
twisted) with $B_{\rm pole}=10^{15}$~G is assumed in this example.
$R$ is the neutron-star radius.
} 
\label{fig:1}       
\end{figure}

To illustrate the drag effect on the $e^\pm$ flow around magnetars, 
consider the following simplified model. 
Suppose that the $e^\pm$ plasma is injected near the star with a given
multiplicity, e.g. $\M=50$, and a given high Lorentz factor, e.g. 
$\gamma_+=100$. The corresponding $\gamma_-$ is determined by 
equation~(\ref{eq:j}).
Suppose that the plasma is illuminated by the blackbody radiation of 
the star of temperature $kT=0.5$~keV and neglect the radiation from the 
magnetosphere itself (see below). 
The steady-state solution for $\gamma_\pm$ established in the outer corona  
is shown in Figure~5. The solution is not sensitive to the 
precise radius of $e^\pm$ injection as long as it is small enough, before 
the plasma enters the drag-dominated region.
The most significant electric field develops in the region where $|\D|\gg 1$.
It is given by $e\Epar\approx -\F(\gamma_+)$, and the 
corresponding longitudinal voltage established in the outer corona
is found by integrating $\F(\gamma_+)$ along the field line,
$e\Phi_e\approx -\int \F(\gamma_+) \,dl$.
Its typical value for the model in Figure~5 is $\sim 10^7$~V.
Flows with lower $\M$ develop stronger electric fields, however in all 
cases of interest ($\M\gg 1$) the drag-induced voltage is below $10^9$~V.

The calculations shown in Figure~5 assume that the plasma is optically 
thin to resonant scattering. This is not so for real magnetars. 
TLK02 showed that the optical depth of a strongly twisted magnetosphere 
is comparable to unity. When the large pair multiplicity $\M$ is taken 
into account, the estimate for the optical depth becomes
\be
\label{eq:tau}
   \tau\sim \frac{\M\psi}{\beta_\pm} \gg 1.
\ee
The large optical depth has two implications. First, there is a lot of 
scattered radiation in the magnetosphere, which is quasi-isotropic. 
This increases the drag and reduces $p_\star$.
Second, there is a self-shielding effect: 
the drag force $\F$ experienced by an electron (or positron) is reduced 
by the factor of $\tau^{-1}$. These two effects will change 
$\gamma_\pm(r,\theta)$ shown in Figure~5.
However, the main feature will persist: the outflow remains to be 
drag-dominated in the equatorial region at $r\sim (10-20)R$.
The initial momentum of the $e^\pm$ flow ejected from the discharge zone 
to the outer corona is taken away by the radiative drag and the $e^\pm$ 
pairs accumulate near the apexes of the closed magnetic field lines. 
The pairs annihilate there. 

The annihilation rate is given by $\dot{N}_{\rm ann}\approx 2\M (I/e)$.
Here $I$ is the electric current through the annihilation region, 
$I\sim (\psi c\mu/4R_1^2)$,~ $\psi$ is the twist amplitude, $R_1$ is the 
inner boundary of the annihilation region, and $\mu$ is the magnetic 
moment of the star.
The corresponding annihilation luminosity is given by
\begin{equation}
L_{\rm ann}=2m_ec^2\M\, \frac{I}{e}
   \sim 4\times 10^{31}\M\,\psi\,\mu_{32}\left(\frac{R_1}{8R}\right)^{-2}
      {\rm ~erg~s}^{-1}.
\end{equation}

\begin{figure}[t]
\hspace{-0.3cm}
\epsfxsize=6.5cm \epsfbox{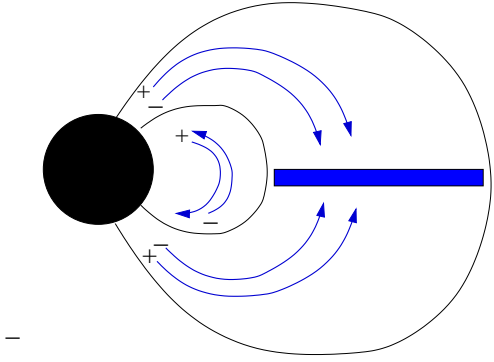}
\caption{
Schematic picture of plasma circulation in the magnetosphere with
surface $B\sim 10^{15}$~G.  Two regions are indicated. (1) ``Inner corona.''
Here $e^\pm$ have a moderate multiplicity $\M\sim 1$. The particles do
not stop in the equatorial plane.  The electric field $\Epar$ ensures that
electrons and positrons circulate in the opposite directions
along the magnetic field lines, maintaining the electric current
demanded by $\nabla\times\bB$. The particles are lost as they reach the
footprints of the field line and continually replenished by pair creation.
(2) ``Outer corona'' --- extended field lines with $\Rmax\gg R$.
  Electrons and positrons are created by the discharge near the star and
  some of them flow outward to the region of weaker $B$.
  Here resonant scattering
  enhances the pair multiplicity, $\M\gg 1$, and decelerates the outflow.
  The $e^\pm$ particles stop at the apexes of magnetic field lines (blue
  region in the equatorial plane), accumulate, and annihilate there.
  The number fluxes of electrons and positrons toward the annihilation
  region differ by a small fraction $\sim\M^{-1}$, so that the outflow
  carries the required electric current $\bj=(c/4\pi)\nabla\times\bB$.
}
\label{fig:1}       
\end{figure}

The proposed picture of the magnetar corona is schematically shown in 
Figure~6. The exact position $R_1$ of the boundary between the two 
regions in Figure~6 has not been calculated and may be inside $8R$. 
Consider, for instance, the plasma flow along the field lines with $\Rmax=5R$.
It experiences scattering with rate that is high enough to generate 
a large $e^\pm$ multiplicity, however, not enough to 
bring the plasma to a stop at the equatorial plane.
The opposite flows in the northern and southern hemispheres meet 
in the equatorial plane and try to penetrate each other.
The two-stream instability then develops and a strong Langmuir turbulence 
is generated, which will inhibit the penetration.
Effectively, the opposite flows collide in the equatorial plane and should 
stop there. This behavior contrasts with the inner corona where the 
counter-streaming $e^-$ and $e^+$ never stop; their circulation is 
maintained by the induced electric field.
The outer corona is different because of its high pair multiplicity $\M$;
then the electric current is organized while
both $e^-$ and $e^+$ {\it outflow} from the star.

The above analysis assumed a steady $e^\pm$ outflow.
Then the current is organized by locking the electrons 
and positrons at two different velocities $\beta_-$ and $\beta_+$.
In reality, the two-fluid flow is prone to two-stream instability that 
generates
strong Langmuir waves and broadens the momentum distribution of $e^\pm$.
The robust generation of plasma waves
gives a mechanism for low-frequency emission from magnetars
(different from drag-induced waves in normal pulsars, cf. 
Lyubarsky \& Petrova 2000). 

The instability has been simulated numerically, and the 
results are described in detail elsewhere (Beloborodov 2010).
The fluctuating electric field $\Epar$ is so strong that it could stop a
particle with Lorentz factor $\gamma_+$ on a short scale, much shorter 
than the free path to resonant scattering.
A strong anomalous resistivity could be expected in this situation.
Surprisingly, the simulations indicate that the effective resistivity 
(and the corresponding voltage) remain low. A complicated time-dependent 
pattern is organized in the phase-space, which allows the charges to 
find small-resistance paths through the waves of $\Epar$ and conduct 
the current at a low net voltage and a low dissipation rate.

\section{Magnetospheric emission}
\label{sec:3}
Various ideas have been proposed for the mechanism of hard X-ray emission
from magnetars
(Thompson \& Beloborodov 2005; Heyl \& Hernquist 2005; BT07; Baring \& 
Harding 2007; Lyubarsky \& Eichler 2008). 

One possible source is the transition layer between the corona and the star. 
The layer can reach temperatures $kT\simgt 100$~keV and produce 
bremsstrahlung emission. 

The other possible source is the extended corona itself. It generates 
nonthermal X-rays via resonant scattering of the thermal X-rays emitted
by the neutron star.
A simplest model assumes that the high-energy spectrum forms via single
scattering by an optically thin flow that interacts with the stellar
radiation. This model was developed for ordinary pulsars 
(e.g. Kardashev, Mitrofanov, \& Novikov 1984; Daugherty \& Harding 1989; 
Sturner 1995; Lyubarsky \& Petrova 2000) and also applied to magnetars 
(Baring \& Harding 2007). It does not, however, explain the magnetar 
spectrum. Several works emphasized that radiative transfer with multiple 
resonant scattering in the closed magnetosphere has a strong impact on
the observed X-ray spectrum (TLK02; Lyutikov \& Gavriil 2006; 
Fernandez \& Thompson 2007; Nobili, Turolla \& Zane 2008; Rea et al. 2008; 
Pavan et al. 2009; Zane et al. 2009).
The key unsettled problem is what parameters of the scattering plasma 
should be assumed in the transfer calculations. Resonant scattering 
depends on the Lorentz factors of the coronal particles and the direction 
of their motion.

One popular assumption is that the corona is filled with 
counter-streaming positive and negative charges with mildly relativistic 
velocities. The scattering in this picture can reproduce  
the 1-10~keV part of the magnetar spectrum  
(e.g. Rea et al. 2008). However its theoretical basis is problematic. 
The keV photons are resonantly scattered at radii $r\sim 10R$ 
where $B\sim 10^{11}-10^{12}$~G.
The plasma strongly interacts with the stellar radiation 
in this region, and the counter-streaming model needs $\Epar$ to push 
charges of one sign toward the star against the radiative drag. 
This electric field accelerates the charges of the opposite sign 
away from the star. In the presence of $e^\pm$ plasma (which is inevitable),
no self-consistent solution exists for the mildly relativistic
counter-streaming model. 

A different picture of plasma circulation is shown in Figure~6.
We argued that in the outer corona 
both $+$ and $-$ charges must outflow from the star 
and annihilate in the equatorial plane of the magnetic dipole. 
A moderate electric field $\Epar$ is generated to maintain 
a small difference $\Delta\beta/\beta\sim \M^{-1}\ll 1$ between the 
average velocities of the positive and negative charges, so that the 
outflow carries the required electric current.
This picture offers a self-consistent solution for the drag-dominated 
electric circuit, which needs to be tested against observations. 
The theoretical X-ray spectrum can be found by solving for the 
transfer of the stellar radiation through the $e^\pm$ outflow.
Then one can compare the theoretical prediction with the data.

The technical difficulty is that the outflow dynamics is coupled to the 
radiation field, and hence the radiative transfer must be solved 
together with the plasma motion. This nonlinear problem is well defined
and can be solved exactly (numerically), however this is a rather 
formidable task.
As a first step, let us consider a simplified model which provides the 
Lorentz factor $\gamma(r,\theta)$ of the scattering medium, motivated
by the results of \S~5.
Given $\gamma(r,\theta)$ it is straightforward to calculate
the transfer of the stellar radiation through the corona.
The result will not depend on the optical depth $\tau$ as 
long as $\tau\gg 1$ (this is a special feature of resonant  scattering).
We use the Monte-Carlo method for the transfer calculations;
it is described elsewhere (Beloborodov 2010).

First, consider the outflow model in Figure~5 and suppose 
$\gamma(r,\theta)=\gamma_+(r,\theta)$ (model~A). The emerging spectrum 
after multiple scattering in this outflow is shown in Figure~7 for
an inclination angle $\theta=\pi/3$.
The model is inconsistent, as Figure~5 assumed that the drag 
force $\F$ is created by the central thermal radiation field, 
neglecting the scattered radiation, which would be appropriate only if
$\tau\ll 1$. The central radiative drag leads to the artificially high 
$\gamma_+$ near the axis where the central radiation is 
beamed along the outflow direction. 

\begin{figure}[t]
  \begin{center}
    \begin{minipage}[t]{0.58\linewidth}
\epsfxsize=7.5cm
 \raisebox{-7.5cm}{\epsfbox{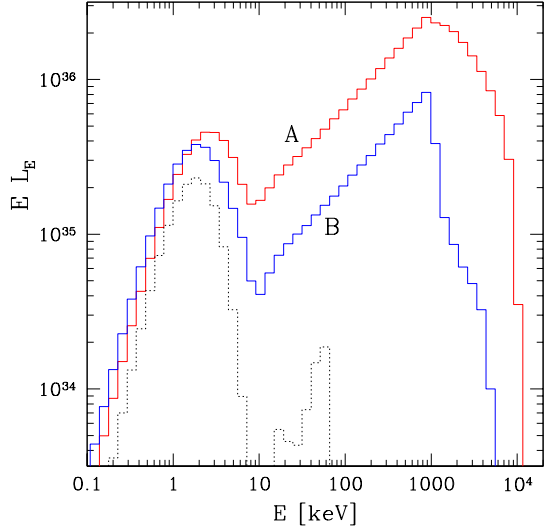}}
    \end{minipage}\hfill
    \begin{minipage}[t]{0.33\linewidth}
\caption{\small
Magnetar spectrum formed by resonant scattering in the $e^\pm$ outflow.
Solid histograms show models A and B (see text). The line of sight is 
chosen at angle $\theta=60^{\rm o}$ with respect to the magnetic axis. 
The star is assumed to emit the blackbody radiation with temperature 
$kT=0.5$~keV. The result of single (first) scattering in Model B is 
shown by the dotted histogram.
}
\end{minipage}
\end{center}
\end{figure}

In reality, $\tau\gg 1$ (even for moderate twists, see eq.~\ref{eq:tau}),
the scattered radiation fills the magnetosphere
and contributes to the drag force, decelerating the outflow near the 
axis. This will lead to small $\gamma(r,\theta)$ at all $\theta$ in the 
outer corona and change the scattered radiation spectrum. To see the trend 
of this change, let us modify the drag force $\F$ in Figure~5.
Let $\alpha$ be the angle between the photons and the outflow velocity. 
In the region where $\cos\alpha>0.5$ for all photons emitted by the star, 
let us set $\cos\alpha=0.5$ when calculating $\F$.
This intrusion makes the radiation field less anisotropic
so that it is never strongly beamed along the outflow. Re-calculating 
$\gamma_\pm(r,\theta)$, we find that they
are reduced compared with those in Figure~5. Now consider the scattering 
medium with the new Lorentz factor $\gamma=\gamma_+$ and calculate the 
radiative transfer through this medium (model~B). The result is shown in 
Figure~7.

Toy models A and B allow one to make some preliminary conclusions.
In both models, the spectrum has two distinct peaks, at a few keV
and near 1~MeV, and a minimum near 10~keV. This shape appears to be a 
robust result of radiative transfer through the $e^\pm$ outflow. 
The result is consistent with the observed spectra of magnetars
(e.g. Kuiper et al. 2008). The photon index $\Gamma$ above 10~keV is close 
to the typical observed $\Gamma\sim 1$.

Multiple scattering is essential for the formation of the high-energy peak. 
For illustration, the dotted histogram in Figure~7 shows 
the spectrum obtained in model~B after single scattering; its high-energy
component is weak. The first scattering of the thermal keV photons occurs 
mainly at radii $r\sim 10R$ where the outflow is slow.
It generates quasi-isotropic radiation with a spectrum that is slightly 
modified from blackbody. A fraction of the scattered photons 
propagate back to the star and approach head-on the relativistic outflow 
in the inner region. The second scattering for such photons dramatically 
boosts their energy, contributing to the high-energy component in the 
observed spectrum. This mechanism will also operate in the exact model 
where the accurate distribution of the Lorentz factor in the outflow is 
found consistently with the radiative transfer. 
In this model, the high-energy luminosity will be explicitly fed by
the initial power of the outflow injected near the star (which is 
determined by the electric current and the discharge voltage).
The radiative drag transforms this power to radiation.

Similar spectra are generated when the outflow is confined
to the j-bundle of moderate magnetic flux $\uf\simgt 0.1$.
The spectrum evolves as the j-bundle shrinks to $\uf\ll 0.1$.
Note also that the high-energy component is anisotropic --- it is 
preferentially beamed along the magnetic dipole axis. This may help 
explain the energy dependence of the pulse profiles in observed 
magnetars (den Hartog et al. 2008a,b). 

The pulse profiles of magnetars are expected to change as the magnetosphere 
untwists and the j-bundle shrinks. Previous discussion of pulse profiles 
assumed that the entire magnetosphere is twisted in a self-similar way 
and has a moderate optical depth $\tau$ proportional to the amplitude of 
the global twist $\psi$ (e.g. Fern\'andez \& Thompson 2007). 
 The calculations presented here suggest a different picture.
The twists of magnetars quickly evolve to 
narrow j-bundles 
(and twists may always be confined to a small part of the magnetosphere).
The plasma in the j-bundle is opaque to resonant scattering even 
for weak twists $\psi\sim 0.1$.
The result of radiative transfer (and the corresponding pulse profile) 
does not directly depend on $\tau$ as long as $\tau\gg 1$. 
It may depend on $\tau$ indirectly, because the deceleration of the $e^\pm$ 
outflow depends on its optical depth. 
Further analysis is needed to explore the implications for the pulse 
profile and to test the model against observations.

A robust result of the radiative transfer (especially in the more 
reasonable model~B) is the sharp break in the spectrum at 
$\Ebr\approx 2m_ec^2\approx 1$~MeV. It is caused by photon absorption 
in the strong magnetic field. The emission of multi-MeV photons occurs
in the region where $B\sim 10^{13}$~G, and they see a large optical depth 
to absorption. Absorption is reduced for photons emitted along the magnetic 
axis, and the emerging high-energy spectrum depends on the line of sight.

Note that somewhat smaller $\Ebr\sim 0.5$~MeV could be possible for 
the emission produced by the transition layer between the corona and star
(Thompson \& Beloborodov 2005; BT07). In that model, $\Ebr$ may be 
controlled by the temperature of the layer rather than absorption in the 
magnetosphere. Observations of $\Ebr$ may help distinguish between the 
two mechanisms of high-energy emission.


\section{Conclusions}
\label{sec:3}
Nonthermal X-ray emission and shrinking hot spots on magnetars 
are associated with magnetospheric activity --- the release of
the energy stored in the twisted closed magnetosphere.
The rate of energy release is proportional to the voltage established 
along the magnetic field lines, which is regulated by the continual 
discharge to $\Phi_e\sim 10^9$~V. This voltage has a pure inductive 
origin, directly related to the rate of untwisting of the magnetic field. 
For axisymmetric configurations, the untwisting evolution is described by 
the electrodynamic equation~(\ref{eq:evol1}).

The untwisting theory predicts the formation of shrinking hot spots on 
``transient magnetars'' whose magnetospheres are temporarily activated 
and gradually relax back 
to the quiescent state. Shrinking hot spots were indeed 
reported in these objects (Fig.~5), and their evolution appears to 
agree with the theoretical expectations.

Self-similar twists (Wolfson 1995; TLK02) do not form in magnetars, 
because the electric currents tend to be quickly removed from field lines 
with moderate apex radius $\Rmax$. Currents have longest lifetime 
on field lines with $\Rmax\gg R$, forming the extended j-bundle. 
An $e^\pm$ outflow of a high multiplicity streams in the j-bundle 
and creates the outer corona around the neutron star. 
The outflow stops at the apexes of the closed magnetic field lines 
and the $e^\pm$ pairs annihilate there.

The $e^\pm$ outflow is opaque to resonant scattering and impacts the 
observed X-ray spectrum. The simulations of radiative transfer through
the outflow suggest a natural explanation for the hard X-ray 
component in magnetar spectra.
 
The strong radiative drag and the imposed electric current lock the 
j-bundle plasma in a peculiar two-fluid state. 
The two-stream instability is then inevitable;
it creates strong Langmuir oscillations, which can convert 
to escaping low-frequency radiation. This mechanism is 
discussed in more detail elsewhere (Beloborodov 2010).

\begin{acknowledgement}
This work was supported by NASA grant NNX-10-AI72G.
\end{acknowledgement}

%
%

\begin{thebibliography}{99.}%
%
%
%
\bibitem{science-journal} 
Baring, M. G., \& Harding, A. K. 2007, Ap\&SS, 308, 109
\bibitem{science-journal} 
Beloborodov, A. M., \& Thompson, C. 2007, ApJ, 657, 967 (BT07)
\bibitem{science-journal} 
Beloborodov, A. M. 2009, ApJ, 703, 1044 (B09)
\bibitem{science-journal} 
Bernardini, F., et al. 2009, A\&A, 498, 195
\bibitem{science-journal} 
Camilo, F. et al. 2007, ApJ, 663, 497
\bibitem{science-journal} 
Daugherty, J. K., \& Harding, A. K. 1989, ApJ, 336, 861
\bibitem{science-journal} 
den Hartog, P.~R., Kuiper, L., \& Hermsen, W. 2008a, A\&A, 489, 263
\bibitem{science-journal} 
den Hartog,~P.~R., Kuiper,~L., Hermsen,~W., Kaspi,~V.~M., Dib,~R.,
Kn\"odlseder,~J., \& Gavriil,~F.~P. 2008b, A\&A, 489, 245
\bibitem{science-journal} 
Dib, R., Kaspi, V. M., \& Gavriil, F. P. 2009, ApJ, 702, 614
\bibitem{science-journal} 
Duncan, R. C., \& Thompson, C. 1992, ApJ, 392, L9
\bibitem{science-journal} 
Durant, M., \& van Kerkwijk, M.~H. 2006, ApJ, 650, 1070
\bibitem{science-journal} 
Enoto, T., et al. 2010a, PASJ, 62, 475
\bibitem{science-journal} 
Enoto, T., et al. 2010b, ApJ, 715, 665
\bibitem{science-journal} 
Esposito, P. et al. 2008, MNRAS, 390, L34
\bibitem{science-journal} 
Esposito, P. et al. 2010, MNRAS, 405, 1787
\bibitem{science-journal} 
Fern\'andez, R., \& Thompson, C. 2007, ApJ, 660, 615
\bibitem{science-journal} 
Gavriil, F.~P. et al. 2008, Science, 319, 1802
\bibitem{science-journal} 
Gavriil, F.~P., Dib, R., \& Kaspi, V. M. 2009, ApJ, submitted (arXiv:0905.1256) 
\bibitem{science-journal} 
Goldreich, P., \& Julian, W.~H.\ 1969, ApJ, 157, 869
\bibitem{science-journal} 
Gonzalez, M. E., Dib, R., Kaspi, V. M., Woods, P. M., Tam, C. R., \& 
Gavriil, F. P. 2010, ApJ, 716, 1345
\bibitem{science-journal} 
Gotthelf, E.~V., \& Halpern, J.~P. 2007, Ap\&SS, 308, 79
\bibitem{science-journal} 
Halpern, J.~P., Gotthelf, E.~V., Reynolds, J., Ransom, S.~M., \& Camilo, F.
2008, ApJ, 676, 1178
\bibitem{science-journal} 
Heyl, J. Sa,. \& Hernquist, L. 2005, MNRAS, 362, 777
\bibitem{science-journal} 
Ibrahim, A.~I., et al. 2004, ApJ, 609, L21
\bibitem{science-journal} 
Israel, G., et al. 2010, MNRAS, in press (arXiv:1006.2950)
\bibitem{science-journal} 
Kardashev, N. S., Mitrofanov, I. G., \& Novikov, I. D. 1984, Sov. Astronomy,
28, 651
\bibitem{science-journal} 
Kouveliotou, C., et al. 2003, ApJ, 596, L79
\bibitem{science-journal} 
Kuiper, L., den Hartog, P.~R., \& Hermsen, W. 2008 (arXiv:0810.4801)
\bibitem{science-journal} 
Kuiper, L., Hermsen, W., den Hartog, P. R., \& Collmar, W. 2006, ApJ, 645, 556
\bibitem{science-journal} 
Low, B.~C. 1986, ApJ, 307, 205
\bibitem{science-journal} 
Lyubarsky, Y., Eichler, D., \& Thompson, C. 2002, ApJ, 580, L69 
\bibitem{science-journal} 
Lyubarsky, Y., \& Eichler, D. 2007, arXiv:0706.3578
\bibitem{science-journal} 
Lyubarsky, Y. E., \& Petrova, S. A. 2000, A\&A, 355, 406
\bibitem{science-journal} 
Lyutikov, M., \& Gavriil, F.~P. 2006, MNRAS, 368, 690
\bibitem{science-journal} 
Lyutikov, M. 2003, MNRAS, 346, 540
\bibitem{science-journal} 
Medin, Z., \& Lai, D. 2010, MNRAS, 406, 1379
\bibitem{science-journal} 
Mereghetti, S., et al. 2006, A\&A, 450, 759
\bibitem{science-journal} 
Mereghetti, S. 2008, Astr. \& Astroph. Rev., 15, 4, 225
\bibitem{science-journal} 
Muno, M. P., et al. 2007, MNRAS, 378, L44
\bibitem{science-journal} 
Nobili, L., Turolla, R., \& Zane, S. 2008, MNRAS, 386, 1527
\bibitem{science-journal} 
Paczy\'nski, B. 1992, Acta Astronomica, 42, 145
\bibitem{science-journal} 
Pavan, L., Turolla, R., Zane, S., \& Nobili, L. 2009, MNRAS, 395, 753
\bibitem{science-journal} 
Perna, R., \& Gotthelf, E.~V. 2008, ApJ, 681, 522
\bibitem{science-journal} 
Perna, R., Heyl, J.~S., Hernquist, L.~E., Juett, A.~M., \& Chakrabarty, D.
2001, ApJ, 557, 18
\bibitem{science-journal} 
Rea,~N., Zane,~S., Turolla,~R., Lyutikov,~M., \& G\"otz,~D. 2008,
\bibitem{science-journal} 
Rea,~N., et al. 2009, MNRAS, 396, 2419
\bibitem{science-journal} 
Sturner, S. J. 1995, ApJ, 446, 292
\bibitem{science-journal} 
Thompson, C., \& Beloborodov, A.~M. 2005, ApJ, 634, 565 
\bibitem{science-journal} 
Thompson, C., \& Duncan, R.~C. 1995, MNRAS, 275, 255
\bibitem{science-journal} 
Thompson, C., \& Duncan, R.~C. 1996, ApJ, 473, 322
\bibitem{science-journal} 
Thompson, C., Lyutikov, M., \& Kulkarni, S.~R. 2002, ApJ, 574, 332 (TLK02)
\bibitem{science-journal} 
Tiengo, A., et al. 2010, ApJ, 710, 227
\bibitem{science-journal} 
Uzdensky, D.~A. 2002, ApJ, 574, 1011
\bibitem{science-journal} 
Wolfson, R. 1995, ApJ, 443, 810
\bibitem{science-journal} 
Wolfson, R., \& Low, B.~C. 1992, ApJ, 391, 353
\bibitem{science-journal} 
Woods, P. M., et al. 2001, ApJ, 552, 748
\bibitem{science-journal} 
Woods, P. M., et al. 2004, ApJ, 605, 378
\bibitem{science-journal} 
Woods, P. M., Kaspi, V. M., Gavriil, F. P., \& Airhart, C. 2010, ApJ, 
submitted (arXiv:1006.5487)
\bibitem{science-journal} 
Woods, P. M., \& Thompson, C. 2006, in Compact Stellar X-Ray Sources,
ed. W.~H.~G. Lewin \& M. van der Klis (Cambridge: Cambridge Univ. Press), 547
\bibitem{science-journal} 
Zane, S., Rea, N., Turolla, R., \& Nobili, L. 2009, MNRAS, 398, 1403
%
%
\end{thebibliography}
%

\end{document}